\documentclass[aps,prd,onecolumn,preprintnumbers,nofootinbib]{revtex4}
\usepackage{amsmath,amssymb,latexsym}
\usepackage{color}

\begin{document}

\title{
Asymptotic flatness at null infinity in arbitrary dimensions
}

\author{Kentaro Tanabe, Shunichiro Kinoshita}
\affiliation{   
Yukawa Institute for Theoretical Physics, Kyoto University, 606-8502 Kyoto, Japan
}
\author{Tetsuya Shiromizu}
\affiliation{
Department of Physics, Kyoto University, 606-8502 Kyoto, Japan
}
\preprint{YITP-11-40} 

\date{\today}

\begin{abstract}
We define the asymptotic flatness and discuss asymptotic symmetry at null infinity 
in arbitrary dimensions using the Bondi coordinates. To define the asymptotic flatness, 
we solve the Einstein equations and look at the asymptotic behavior of gravitational fields. 
Then we show the 
asymptotic symmetry and the Bondi mass loss law with the well-defined definition. 
\end{abstract}

\maketitle

 \section{Introduction}
 
 Recently inspired by string theory, the systematic investigation of gravitational theory 
in higher dimensional spacetimes becomes more important. Indeed there are many things to be studied. 
Asymptotic structure of higher dimensional spacetimes is one of them. Asymptotically flat spacetimes 
have spatial and null infinities. While the asymptotic 
structure at spatial infinity has been investigated in arbitrary  dimensions 
\cite{Ashtekar:1978zz, Tanabe:2009xb}, the studies on null infinity have been done only in 
even dimensions \cite{Geroch:1977jn, Bondi:1962px, Sachs:1962wk, Sachs:1962, Hollands:2003ie, 
Hollands:2003xp, Ishibashi:2007kb} and five dimensions \cite{Tanabe:2009va, Tanabe:2010rm}. 
 
 In four dimensions, asymptotic structure at null infinity was investigated in two ways. One is 
based on the Bondi coordinates \cite{Bondi:1962px, Sachs:1962wk} and the another is based on 
the conformal embedding \cite{Geroch:1977jn,Penrose:1962ij}. In the latter, we introduce the 
conformal factor $\Omega\sim 1/r$ and we can 
study the behavior of gravitational fields near null infinity in the conformally transformed spacetime. 
This method can be extended to 
higher dimensions, but even dimensions only \cite{Hollands:2003ie, Hollands:2003xp, Ishibashi:2007kb}. The 
reason why {\it even dimensions} is as follows. In $n$-dimensional spacetimes, 
gravitational fields behave like $\sim 1/r^{n/2-1} \sim \Omega^{n/2-1}$ and then 
we cannot suppose the smoothness of gravitational fields at null infinity due to the power of the half-integer 
in odd dimensions. Thus the conformal embedding method will not be useful for 
the investigation of the asymptotic structure at null infinity in any dimensions.

Instead of the conformal embedding method, we could safely define asymptotic flatness and study the 
asymptotic structure at null infinity in {\it five} dimensions by using the Bondi coordinates 
\cite{Tanabe:2009va}. Therein one must solve the Einstein equations and determines the asymptotic 
behavior of gravitational fields which gives us a natural definition of asymptotic flatness at 
null infinity. We can show the asymptotic symmetry and the finiteness of the 
Bondi mass in five dimensions. The purpose of this paper is the extension of this work to 
arbitrary dimensions. For simplicity, we will consider the spacetimes satisfying the vacuum 
Einstein equation in higher dimensions. But, it is easy to extend our current work into nonvacuum 
cases that matters rapidly decay near null infinity. 

 The remaining part of this paper is organized as follows. In Sec.~\ref{sec:Bondi_coodinates}, we introduce the 
Bondi coordinates in $n$-dimensional spacetimes and write down the Einstein equations in the 
language of the ADM formalism. In Sec.~\ref{sec:asymptotic_flatness}, 
the Einstein equations will be solved explicitly and asymptotic flatness is defined by 
asymptotic behaviors at null infinity. We also define the Bondi mass and show its finiteness 
and the Bondi mass loss law. In Sec.~\ref{sec:asymptotic_symmetry}, we shall study the asymptotic symmetry. In Sec.~\ref{sec:summary} we will 
summarize our work and discuss our future work. In 
Appendix~\ref{app:decomposition}, we give the formulae of the ADM 
decomposition which will be used in Sec.~\ref{sec:asymptotic_flatness} and in Appendix~\ref{app:derivations} we show the detail derivations of 
some equations.          

\section{Bondi coordinates and ADM decomposition}
\label{sec:Bondi_coodinates}

 We introduce the Bondi coordinates in $n$-dimensional spacetimes. In the Bondi coordinates 
the metric can be written as 
%============<Equation>=============%
%
\begin{equation}
   \begin{aligned}
    \mathrm ds^2 =& - Ae^{B} \mathrm du^2 - 2e^B \mathrm du\mathrm dr 
    + \gamma_{IJ}(\mathrm dx^I + C^I\mathrm du)
    (\mathrm dx^J + C^J\mathrm du),
\end{aligned}
\end{equation}
%
%==================================%
where $x^{a}=(u, r, x^I)$ denote the retarded time, radial coordinate and angular coordinates, 
respectively. We also impose the gauge condition as
%============<Equation>=============%
%
\begin{eqnarray}
\sqrt{\det\gamma_{IJ}}\,=\,r^{n-2}\omega_{n-2} \label{gauge},
\end{eqnarray}
%
%==================================%
where $\omega_{n-2}$ is the volume element on the unit $(n-2)$-dimensional sphere. 
In the Bondi coordinates, null infinity is located at $r=\infty$. 
 
We perform the ADM decomposition with respect to the $r$-constant surfaces (see Appendix A). 
The metric is rewritten as 
%============<Equation>=============%
%
\begin{eqnarray}
 \mathrm ds^2 = N^2 \mathrm dr^2
   + q_{\mu\nu}
   (\mathrm dx^\mu + N^\mu\mathrm dr)(\mathrm dx^\nu + N^\nu\mathrm dr),
\end{eqnarray} 
%
%==================================% 
where 
%============<Equation>=============%
%
\begin{eqnarray}
& & N^2 = \frac{e^B}{A},\\
& &  N^u = \frac{1}{A}, \\
& & N^I = - \frac{C^I}{A},
\end{eqnarray}
% 
%==================================%
and the induced metric of the $r$-constant surface 
%============<Equation>=============%
%
\begin{eqnarray}
 q_{\mu\nu} = 
   \begin{pmatrix}
    - Ae^B + C^IC_I & C_J\\
    C_I & \gamma_{IJ}
   \end{pmatrix}.
\end{eqnarray}
%
%==================================%
Note that the capital Latin indices $I,J,\cdots$ and  the Greek indices $\mu,\nu,\cdots$ 
are raised and lowered using $\gamma_{IJ}$ and $q_{\mu\nu}$, respectively. 
The unit normal vector to the $r$-constant surface is given by 
$n_a = N (\mathrm dr)_a$ and $n^a = N^{-1}(\partial_r - N^\mu\partial_\mu)^a$. 
The extrinsic curvature of $r$-constant surface is defined as usual 
%============<Equation>=============%
%
\begin{eqnarray}
 K_{\mu\nu} = \frac{1}{2}\mathcal L_n q_{\mu\nu}
   = \frac{1}{2N}(\partial_r q_{\mu\nu} - \mathcal D_\mu N_\nu - \mathcal D_\nu N_\mu),
\end{eqnarray}
%
%==================================% 
where $\mathcal D_\mu$ denotes the covariant derivative associated with $q_{\mu\nu}$.

The induced metric on the $r$-constant surface is rewritten as 
%============<Equation>=============%
%
\begin{eqnarray}
  q_{\mu\nu}\mathrm dx^\mu\mathrm dx^\nu = 
   - \alpha^2 \mathrm du^2 + \gamma_{IJ}
   (\mathrm dx^I + \beta^I\mathrm du)
   (\mathrm dx^J + \beta^J\mathrm du),
\end{eqnarray}
%
%==================================% 
where 
%============<Equation>=============%
%
\begin{eqnarray}
  \alpha^2 = Ae^B, \quad \beta^I = C^I.
\end{eqnarray}
%
%==================================% 

The timelike unit normal vector to the $u$-constant surface is written as 
$u_a = - \alpha (\mathrm du)_a$
\footnote{
This expression is valid on the induced manifold determined
by the $r$-constant surface. For the whole spacetime manifold we should
write it as $u_a = - \alpha (\mathrm du)_a - N (\mathrm dr)_a$
} and 
$u^a = \alpha^{-1}(\partial_u - \beta^I\partial_I)^a$.
The extrinsic curvature of $u$-constant surface becomes 
%============<Equation>=============%
%
\begin{eqnarray}
  k_{IJ} = \frac{1}{2}\mathcal L_u \gamma_{IJ}
   = \frac{1}{2\alpha}(\partial_u\gamma_{IJ}
   - D_I\beta_J - D_J\beta_I),
\end{eqnarray}
%
%==================================%
where $D_I$ denotes the covariant derivative associated with 
$\gamma_{IJ}$. We have $N^a = \alpha A^{-1} u^a = Nu^a$ from the definitions of $u^a$ and $N^a$. 
Also we have $n^a = N^{-1}(\partial_r)^a - u^a$. 

For later convenience we define the 
following projected quantities on the $(n-2)$-dimensional space as 
%============<Equation>=============%
%
\begin{eqnarray}
  \theta \equiv & K_{\mu\nu}u^\mu u^\nu = - \frac{1}{N}\partial_r(\log\alpha) 
   %+ u^\mu\nabla_\mu \log N,\\
   + \mathcal L_u \log N,\\
   \rho^I \equiv & K^I{}_{\mu}u^\mu = \frac{1}{2N\alpha}\partial_r \beta^I
   + \frac{1}{2} D^I \log\frac{N}{\alpha}, \\
   \sigma_{IJ} \equiv & K_{KL}\gamma_I{}^K\gamma_J{}^L =
   \frac{1}{2N}\partial_r \gamma_{IJ} - k_{IJ},
\end{eqnarray}
%
%==================================% 
 where $\rho_\mu u^\mu = \sigma_{\mu\nu}u^\mu = 0$. Using of them, 
$K_{\mu\nu}$ is expressed by 
%============<Equation>=============%
%
\begin{eqnarray}
  K_{\mu\nu} = \theta u_\mu u_\nu - 2\rho_{(\mu}u_{\nu)} 
   + \sigma_{\mu\nu}.
\end{eqnarray}
%
%==================================%

\subsection{Decomposition of Ricci tensor}

The vacuum Einstein equation is $\hat R_{ab}=0$. 
Let us decompose the $n$-dimensional Ricci tensor $\hat R_{ab}$ into the 
quantities on the $(n-2)$-dimensional space:
%============<Equation>=============%
%
\begin{equation}
   \begin{aligned}
   \hat R_{ab}n^an^b %=& - \mathcal L_n K - K_{\mu\nu}K^{\mu\nu}
    %- \frac{1}{N}\nabla^2N\\
    =& \frac{1}{N}(\theta - \sigma)'
    - \mathcal L_u(\theta - \sigma)
    - \theta^2 + 2\rho^I\rho_I
    - \sigma_{IJ}\sigma^{IJ}
    + \frac{1}{N}\mathcal L_u\mathcal L_u N
    + k \mathcal L_u \log N\\
    &- \frac{1}{N} D^2 N
    - D_I\log\alpha D^I\log N, 
   \end{aligned}
   \label{eq:Ricci_nn}
\end{equation}
%
%==================================%
%============<Equation>=============%
%
\begin{equation}
   \begin{aligned}
    \hat R_{ab}n^au^b %=& u^\nu \nabla^\mu (K_{\mu\nu} - q_{\mu\nu}K)\\
    =& - \mathcal L_u \sigma - \theta k - k_{IJ}\sigma^{IJ}
    + D_I\rho^I + 2\rho^I D_I\log\alpha,
   \end{aligned}
   \label{eq:Ricci_nu}
\end{equation} 
%
%==================================%
%============<Equation>=============%
%
\begin{equation}
   \begin{aligned}
    \hat R_{ab}u^au^b %=& 
    %{}^{(q)}R_{\mu\nu}u^\mu u^\nu - u^\mu u^\nu\mathcal L_n K_{\mu\nu}
    %- KK_{\mu\nu}u^\mu u^\nu + 2 K_{\mu \alpha}K_\nu{}^\alpha u^\mu u^\nu
    %- u^\mu u^\nu\frac{1}{N}\nabla_\mu \nabla_\nu N
    %\\
    =& - \frac{1}{N}\theta' + \theta^2 - \theta\sigma - 2\rho^I\rho_I
    + 2\rho^I D_I\log\frac{N}{\alpha}
    + D^I\log N D_I\log\alpha +
    \frac{1}{\alpha} D^2\alpha\\
    & - \mathcal L_u k - k_{IJ}k^{IJ} + \mathcal L_u\theta
    - \frac{1}{N}\mathcal L_u\mathcal L_u N,
   \end{aligned}
   \label{eq:Ricci_uu}
\end{equation} 
%
%==================================%
%============<Equation>=============%
%
\begin{equation}
   \begin{aligned}
    \hat R_{ab}n^a \gamma^{bI} %=&
    %\gamma^{IJ} \nabla^\mu (K_{\mu J} - q_{\mu J}K)\\
    =& \theta D^I\log\alpha 
    - 2\rho^J k_J{}^I - \rho^I k 
    + \sigma^{IJ} D_J\log\alpha
    + D_J \sigma^{IJ} - D^I\sigma
    + D^I\theta
    - \frac{1}{\alpha}[\dot{(\rho^I)} - \mathcal L_\beta \rho^I],
   \end{aligned}
   \label{eq:Ricci_ng}
\end{equation}
%
%==================================%  
%============<Equation>=============%
%
\begin{equation}
   \begin{aligned}
    \hat R_{ab}u^a\gamma^{bI} %=& 
    %{}^{(q)}R_{\mu\nu}u^\mu \gamma^{\nu I}
    %- u^\mu \gamma^{\nu I}\mathcal L_n K_{\mu\nu}
    %- KK_{\mu\nu}u^\mu \gamma^{\nu I}
    %+ 2 K_{\mu\alpha}K_\nu{}^\alpha u^\mu \gamma^{\nu I}
    %- u^\mu \gamma^{\nu I}\frac{1}{N}\nabla_\mu\nabla_\nu N\\
    =& D_J (k^{IJ} - \gamma^{IJ}k)
    - \sigma \rho^I
    - 2\rho_J \sigma^{IJ}
    - \frac{1}{N} D^I\mathcal L_u N
    + k^{IJ} D_J\log N\\
    & - \frac{1}{N}(\rho^I)' %- 4\sigma^{IJ}\rho_J
    %+ \theta\rho^I
    + \theta D^I\log\frac{N}{\alpha}
    + \sigma^{IJ} D_J\log\frac{N}{\alpha}
    + \frac{1}{\alpha}[\dot{(\rho^I)} - \mathcal L_\beta \rho^I],
   \end{aligned}
   \label{eq:Ricci_ug}
\end{equation}
%
%==================================%  
%============<Equation>=============%
%
\begin{equation}
   \begin{aligned}
    \hat R_{ab}\gamma^{ab} =&
    - \frac{1}{N}\sigma' + \mathcal L_u \sigma
    + \theta\sigma - \sigma^2 - 2\rho^I\rho_I
    + 2\rho^I D_I \log \frac{N}{\alpha}
    \\
    &+ \mathcal L_u k + k^2
    - \frac{1}{\alpha} D^2\alpha
    - \frac{1}{N} D^2 N
    + k\mathcal L_u\log N
    + {}^{(\gamma)}R
   \end{aligned}
   \label{eq:Ricci_gtrace}
\end{equation}
%
%==================================%
and
%============<Equation>=============%
%
\begin{equation}
   \begin{aligned}
    \hat R_{ab}\gamma_I{}^a\gamma_J{}^b =&
    - \frac{1}{N}\sigma_{IJ}' + \frac{1}{\alpha}\dot\sigma_{IJ}
    - \frac{1}{\alpha}\mathcal L_\beta \sigma_{IJ}
    + \rho_I D_J\log\frac{N}{\alpha}
    + \rho_J D_I\log\frac{N}{\alpha}
    \\
    & + (\theta-\sigma)\sigma_{IJ} - 2\rho_I\rho_J
    + 2 \sigma_{IK}\sigma_J{}^K - \frac{1}{N} D_I D_J N
    - \frac{1}{\alpha} D_I D_J \alpha
    \\
    &+ \mathcal L_u k_{IJ} + kk_{IJ}
    - 2k_{IK}k_J{}^K + k_{IJ}\mathcal L_u\log N
    + {}^{(\gamma)}R_{IJ},
   \end{aligned}
   \label{eq:Ricci_gg}
\end{equation}
%
%==================================%
  %  \footnote{$
  %  \gamma^{\mu\nu} \mathcal L_n K_{\mu\nu} = 
  %  \mathcal L_n \sigma + 2 \sigma^{IJ}\sigma_{IJ}
  %  - 2\rho_I D^I\log \frac{N}{\alpha}
  %  $}
  %  \footnote{$
  %  u^\mu u^\nu \mathcal L_n K_{\mu\nu} = 
  %  \frac{1}{N}\theta' - \mathcal L_u \theta - 2\theta^2
  %  + 4\rho^I\rho_I - 2\rho^I D_I\log \frac{N}{\alpha}
  %  $}
  %  \footnote{$
  %  q^{\mu\nu}\mathcal L_n K_{\mu\nu} = 
  %  \frac{1}{N}(-\theta + \sigma)' - \mathcal L_u (-\theta + \sigma)
  %  + 2\theta^2 - 4\rho^I\rho_I + 2\sigma_{IJ}\sigma^{IJ}
  %  $}
  %  \footnote{$
  %  \gamma_I{}^\mu\gamma_J{}^\nu \mathcal L_n K_{\mu\nu} = 
  %  \frac{1}{N}\sigma_{IJ}' - \frac{1}{\alpha}\dot\sigma_{IJ}
  %  + \frac{1}{\alpha}\mathcal L_\beta \sigma_{IJ}
  %  - \rho_I D_J\log\frac{N}{\alpha}
  %  - \rho_J D_I\log\frac{N}{\alpha}
  %  $}
where the prime and the dot respectively denote $\partial_r$ and $\partial_u$, and ${}^{(\gamma)}R_{IJ}$ denotes the Ricci tensors 
with respect to $\gamma_{IJ}$. In the aboves, we have used the following 
equations
%============<Equation>=============%
%
\begin{equation}
   \begin{aligned}
    \gamma^{\mu\nu} \mathcal L_n K_{\mu\nu} =& 
    \mathcal L_n \sigma + 2 \sigma^{IJ}\sigma_{IJ}
    - 2\rho_I D^I\log \frac{N}{\alpha},\\
    u^\mu u^\nu \mathcal L_n K_{\mu\nu} =& 
    \frac{1}{N}\theta' - \mathcal L_u \theta - 2\theta^2
    + 4\rho^I\rho_I - 2\rho^I D_I\log \frac{N}{\alpha},\\
    q^{\mu\nu}\mathcal L_n K_{\mu\nu} =& 
    \frac{1}{N}(-\theta + \sigma)' - \mathcal L_u (-\theta + \sigma)
    + 2\theta^2 - 4\rho^I\rho_I + 2\sigma_{IJ}\sigma^{IJ},\\
    \gamma_I{}^\mu\gamma_J{}^\nu \mathcal L_n K_{\mu\nu} =& 
    \frac{1}{N}\sigma_{IJ}' - \frac{1}{\alpha}\dot\sigma_{IJ}
    + \frac{1}{\alpha}\mathcal L_\beta \sigma_{IJ}
    - \rho_I D_J\log\frac{N}{\alpha}
    - \rho_J D_I\log\frac{N}{\alpha}.
   \end{aligned}
\end{equation}
%
%==================================%

\section{Asymptotic flatness at null infinity and Bondi mass}
\label{sec:asymptotic_flatness}
 
 In this section, we first solve the Einstein equations near null infinity and examine the 
asymptotic behaviors of the gravitational fields. Then these considerations give us the 
natural definition of the asymptotic flatness at null infinity. We also give the definition 
of the Bondi mass and momenta and then show its finiteness. In the following, we write 
$\gamma_{IJ}$ as $\gamma_{IJ}=r^{2}h_{IJ}$ and the indices $I, J$ are raised and lowered 
by $h_{IJ}$.

\subsection{Constraint equations}
  
Components of the vacuum Einstein equations $\hat{R}_{ra}=0$ and $\hat{R}_{ab}\gamma^{ab}=0$ 
are the constraint equations which do not contain $u$-derivative terms in the current 
coordinate system. Then, once we solve the equations $\hat{R}_{ra}=0$ on the initial 
$u$-constant surface, $\hat{R}_{ra}=0$ always hold in any $u$-constant surfaces.  
  
After direct calculations $\hat R_{rr} = 0$ becomes  
%============<Equation>=============%
%
\begin{equation}
   B' = \frac{r}{4(n-2)}h_{IJ}' h_{KL}' h^{IK}h^{JL} \label{B-eq},
\end{equation}
%
%==================================%
where the prime stands for the $r$-derivative. From $\hat R_{ab}\gamma^{ab} = 0$ we have 
%============<Equation>=============%
%
\begin{equation}
   \begin{split}
    (n-2)\frac{(r^{n-3}A)'}{r^{n-2}}
    = - \nabla_I {C^I}' - \frac{2(n-2)}{r}\nabla_I C^I
    - \frac{r^2e^{-B}}{2}h_{IJ}{C^I}'{C^J}'\\
    - \frac{e^{B}}{2r^2}h^{IJ}\nabla_I B \nabla_J B
    - \frac{e^{B}}{r^2}\nabla_I(h^{IJ}\nabla_J B)
    + \frac{e^{B}}{r^2}{}^{(h)}R,
\end{split}
\label{A-eq}\end{equation}
%
%==================================%
%where $\nabla_{I}$ and ${}^{(h)}\nabla$ denote the covariant derivative with respect to 
%the metric of the unit $(n-2)$-sphere $\omega_{IJ}$ and $h_{IJ}$, respectively. 
and from $\hat R_{rJ}\gamma^{IJ} = 0$ we have 
%============<Equation>=============%
%
\begin{equation}
   \frac{1}{r^{n-2}}(r^n e^{-B}h_{IJ}{C^J}')'
    = - \nabla_I B' + \frac{n-2}{r}\nabla_I B
    + {}^{(h)}\nabla^J h_{IJ}' \label{C-eq},
\end{equation}
%
%==================================%
where $\nabla_{I}$ and ${}^{(h)}\nabla_{I}$ denote the covariant derivative with respect to 
the metric of the unit $(n-2)$-sphere $\omega_{IJ}$ and $h_{IJ}$, respectively. 
${}^{(h)}R$ is the Ricci scalar with respect to $h_{IJ}$.

Once $h_{IJ}$ are given on the initial $u$-constant surface, 
the other metric functions $A, B, C^{I}$ are automatically determined through the 
above equations on the initial $u$-constant surface. 
As seen later, it turns out that $h_{IJ}$ contains the 
degree of freedom of gravitational waves in $n$-dimensional spacetimes. 

We would suppose that $h_{IJ}$ behaves near null infinity as follows
%============<Equation>=============%
%
\begin{eqnarray}
  h_{IJ}\,=\,\omega_{IJ}+\sum_{k\geq 0}h_{IJ}^{(k+1)}r^{-(n/2+k-1)} 
          =\omega_{IJ}+O(r^{-(n/2-1)})\label{bc},
\end{eqnarray}
%
%==================================%
where the summation is taken over $k\in \bf{Z}$ for even dimensions and 
$2k\in \bf{Z}$ for odd dimensions. 
This comes from the fact that $h_{IJ}$ corresponds to gravitational waves. 
By the gauge conditions of Eq.~(\ref{gauge}), 
$h_{IJ}^{(k+1)}$ should be traceless for $k<n/2-1$. 
Notice that we required the fall-off condition which is expected through 
the asymptotic behaviors of linear perturbations around the Minkowski spacetime. 
If the fall-off of $h_{IJ}$ would be $O(r^k)$ for $k>-(n/2-1)$, the nonlinear feature would 
appear in the leading orders and then the Bondi mass will diverge. 
As shown later, the 
condition Eq.~(\ref{bc}) corresponds to the outgoing boundary condition at null infinity.

By Eqs.~(\ref{B-eq}) and (\ref{bc}), we can see that $B$ behaves near null infinity as 
%============<Equation>=============%
%
\begin{eqnarray}
  B\,=\,B^{(1)}r^{-(n-2)} +O(r^{-(n-3/2)}), 
\end{eqnarray}
%
%==================================% 
where 
%============<Equation>=============%
%
\begin{eqnarray}
  B^{(1)}\,=\,-\frac{1}{16}\omega^{IK}\omega^{JL}h^{(1)}_{IJ}h^{(1)}_{KL}.
\end{eqnarray}
%
%==================================%
  
Substituting Eq.~(\ref{C-eq}) into Eq.~(\ref{bc}), we find that $C^{I}$ should behave
%============<Equation>=============%
%
\begin{eqnarray}
  C^{I}\,=\,\sum_{k=0}^{k<n/2-1}C^{(k+1)I}r^{-(n/2+k)} + J^{I}(u,x^{I})r^{-(n-1)} +O(r^{-(n-1/2)}),
\end{eqnarray}
%
%==================================%
where
%============<Equation>=============%
%
\begin{gather}
  C^{(k+1)I}\,=\,\frac{2(n+2k-2)}{(n+2k)(n-2k-2)}\nabla_{J}h^{(k+1)IJ}
  \label{Csol}.
\end{gather}
%
%==================================%
$J^{I}(u,x^J)$ is the integration function in the $r$-integration. 
It corresponds to the angular momentum at null infinity.
 From the terms of the order of $O(r^{-(n-1)})$ terms in Eq.~(\ref{C-eq}), 
we obtain the constraint conditions on $h^{(n/2)}_{IJ}$ as
%============<Equation>=============%
%
\begin{eqnarray}
  \nabla^{J}h^{(n/2)}_{IJ}\,=\,2\nabla_{I}B^{(1)}.
\end{eqnarray}
%
%==================================%  
  
Substituting Eq.~(\ref{A-eq}) into Eq.~(\ref{bc}), we find that $A$ should behave as 
%============<Equation>=============%
%
\begin{eqnarray}
  A\,=\,1+\sum_{k=0}^{k<n/2-2}A^{(k+1)}r^{-(n/2+k-1)} -m(u,x^{I})r^{-(n-3)} +O(r^{-(n-5/2)}),
\end{eqnarray}
%
%==================================%
where
%============<Equation>=============%
%
\begin{eqnarray}
  A^{(k+1)}\,=\,-\frac{2(n+2k-4)}{(n-2k-4)(n+2k-2)}\nabla^{I}C^{(k+1)}_{I}
           \,=\,-\frac{4(n+2k-4)}{(n+2k)(n-2k-2)(n-2k-4)}\nabla^{I}
                \nabla^{J}h^{(k+1)}_{IJ}.
                \label{Asol}
\end{eqnarray}
%
%==================================%
$m (u,x^I)$ is the integration function in the $r$-integration. It corresponds 
to the energy and momentum at null infinity. From the terms of the order of $O(r^{-(n-1)})$ in 
Eq.~(\ref{A-eq}), we obtain the constraint conditions on $h_{IJ}^{(n/2-1)}$ as
%============<Equation>=============%
%
\begin{eqnarray}
  \nabla^{I}C^{(n/2-1)}_{I}\,=\,\nabla^{I}\nabla^{J}h^{(n/2-1)}_{IJ}\,=\,0.
\label{const}
\end{eqnarray}
%
%==================================%
  
To be summarized, if we impose the boundary condition on $h_{IJ}$ as 
Eq.~(\ref{bc}), the behavior of other metric functions $A, B$ and $C^{I}$ 
near null infinity are determined. We will regard asymptotic behaviors as the 
definition of asymptotic flatness at null infinity in $n$-dimensional spacetimes as
%============<Equation>=============%
%
\begin{eqnarray}
   h_{IJ}\,=\,\omega_{IJ} +O(r^{-(n/2-1)}) \,,\,
   A\,=\,1+O(r^{-(n/2-1)}) \,,\,
   B\,=\,O(r^{-(n-2)}) \,,\,
   C^{I}\,=\,O(r^{-(n/2)}).
\end{eqnarray}
%
%==================================%

\subsection{Bondi mass}
  
Next we define the Bondi mass at null infinity in $n$-dimensional spacetimes. 
Since $g_{uu}$ is expanded near null infinity as
%============<Equation>=============%
%
\begin{eqnarray}
   g_{uu}\,=\,-1-\sum_{k=0}^{k<n/2-2}\frac{A^{(k+1)}}{r^{n/2+k-1}} +\frac{m(u,x^{I})}{r^{n-3}} 
   +O(r^{-(n-5/2)}),
\end{eqnarray}
%
%==================================%
we define the Bondi mass $M_{\text{Bondi}}(u)$ and momentum $M^{i}_{\text{Bondi}}(u)$ as
%============<Equation>=============%
%
\begin{eqnarray}
    M_{\text{Bondi}}(u) &\equiv & \frac{n-2}{16\pi}\int_{S^{n-2}}m d\Omega, \\ 
    M^{i}_{\text{Bondi}}(u) &\equiv & \frac{n-2}{16\pi}\int_{S^{n-2}}m\hat{x}^{i} d\Omega, 
\end{eqnarray}
%
%==================================%
respectively. 
$\hat{x}^{i}$ is the unit normal vector to the $(n-2)$-dimensional sphere which satisfies 
$\nabla_{I}\nabla_{J}\hat{x}^{i}+\omega_{IJ}\hat{x}^{i}\,=\,0$. 
Thus each component of $\hat{x}^{i}$ is described by linear combination of the $l=1$ modes of the scalar
harmonics on $S^{n-2}$.
The Bondi 
energy-momentum is defined as $M^{a}_{\text{Bondi}}=(M_{\text{Bondi}},M^{i}_{\text{Bondi}})$.
  
In the conformal method~\cite{Hollands:2003ie, Hollands:2003xp, Ishibashi:2007kb}, the Bondi mass is defined as 
$M_{\text{Bondi}}\sim\int_{S^{n-2}}r^{n-1}\hat{C}_{urur}
  d\Omega\sim\int_{S^{n-2}}r^{n-1}\partial^{2}_{r}g_{uu}d\Omega$, 
where $\hat{C}_{abcd}$ is the $n$-dimensional Weyl tensor. 
At first glance it seems to diverge at null infinity because 
$r^{n-1}\partial^{2}_{r}g_{uu}\sim r^{n/2-2}A^{(1)}$
(it is shown that the Bondi 
mass is finite via an indirect argument in the conformal method \cite{Ishibashi:2007kb}.). However, 
since $A^{(k+1)}$ can be written as 
$A^{(k+1)}\propto \nabla^{I}\nabla^{J}h^{(k+1)}_{IJ}$ (see Eq.~(\ref{Asol})), 
$A^{(k+1)}$ has no contribution to the mass 
and momentum at null infinity for $k<n/2-2$ as
%============<Equation>=============%
%
\begin{eqnarray}
  \int_{S^{n-2}}A^{(k+1)}d\Omega\,=\,\int_{S^{n-2}}\hat{x}^{i}A^{(k+1)}d\Omega\,=\,0.
\end{eqnarray}
%
%==================================%
Thus the Einstein equations guarantee the finiteness of the Bondi mass and momentum regardless of the 
dimension. 
  
\subsection{Evolution equations}
  
The remaining components of the Einstein equation describe the evolution equations 
of gravitational fields. The equation 
$\hat{R}_{ab}\gamma^{a}_{~I}\gamma^{b}_{~J}=0$ 
represents the evolutions of $h_{IJ}$. Indeed near null infinity 
we can obtain for $0\le k<n/2-1$, 
%============<Equation>=============%
%
\begin{eqnarray}
  (k+1)\dot{h}_{IJ}^{(k+2)}&=&-\frac{1}{2}(n-2k-4)A^{(k+1)}\omega_{IJ} 
   +\frac{1}{8}[ n^{2}-6n-(4k^{2}+4k-16)] h^{(k+1)}_{IJ}\notag \\
   &&~~~
   +\frac{1}{2}( -\nabla^{2}h^{(k+1)}_{IJ}+2\nabla_{(I}\nabla^{K}h^{(k+1)}_{J)K})
   -\frac{1}{2}(n-2k-4)\nabla_{(I}C^{(k+1)}_{J)}
   -\nabla^{K}C^{(k+1)}_{K}\omega_{IJ} , 
   \label{h-evo}
\end{eqnarray}
%
%==================================%
where the dot denotes the $u$-derivative. Note that 
the evolutions of $h^{(1)}_{IJ}$ cannot be determined from the above equation. 
$\dot{h}^{(1)}_{IJ}$ are free functions on the initial $u$-constant surface. 
Contracting Eq.~(\ref{h-evo}) with $\nabla^{I}\nabla^{J}$ and using Eq.~(\ref{Asol}), 
we can obtain the evolution equations of $A^{(k+1)}$ as
%============<Equation>=============%
%
\begin{eqnarray}
   \dot{A}^{(k+2)}\,=\,-\frac{n+2k-2}{2(k+1)(n+2k+2)}\nabla^{2}A^{(k+1)}
   +\frac{(n+2k-2)^{2}(n-2k-4)}{8(k+1)(n+2k+2)}A^{(k+1)} 
   \label{A-evo}.
\end{eqnarray}
%
%==================================%
  
From $\hat{R}_{ab}u^{a}u^{b}=0$, we can obtain the evolution equation of $m(u,x^{I})$ as
%============<Equation>=============%
%
\begin{eqnarray}
   \dot{m}\,=\,-\frac{1}{2(n-2)}\dot{h}_{IJ}^{(1)}\dot{h}^{(1)IJ}
   +\frac{n-5}{n-2}\nabla^{I}C^{(n/2-2)}_{I}
   +\frac{1}{n-2}\nabla^{2}A^{(n/2-2)}.
\end{eqnarray}
%
%==================================%
Integrating this equation over the unit $(n-2)$-sphere, we can obtain the Bondi mass loss law as
%============<Equation>=============%
%
\begin{eqnarray}
   \frac{d}{du}M_{\text{Bondi}}\,=\,-\frac{1}{32\pi}\int_{S^{n-2}}
   \dot{h}_{IJ}^{(1)}\dot{h}^{(1)IJ} d\Omega \leq 0 
   \label{BML}.
\end{eqnarray}
%
%==================================%
Thus, the Bondi mass always decreases by gravitational waves and 
this justifies that our boundary conditions of Eq.~(\ref{bc}) correspond to the 
outgoing boundary condition at null infinity.

\section{Asymptotic symmetry}
\label{sec:asymptotic_symmetry}
 
In this section we discuss the asymptotic symmetry at null infinity. We also confirm the 
Poincar\'e covariance of the Bondi mass and momentum.

  \subsection{Asymptotic symmetry}
 
  Asymptotic symmetry is defined to be the transformation group which preserves 
the asymptotic structure at null infinity. The variations of asymptotic form of metric 
at null infinity are given by 
%============<Equation>=============%
%
\begin{gather}
  \delta g_{rr}\,=\,0\,,\,\delta g_{rI}\,=\,0\,,\,g^{IJ}\delta g_{IJ}\,=\,0 \label{AS1}\\
  \delta g_{uu}\,=\,O(r^{-(n/2-1)})\,,\,\delta g_{uI}\,=\,O(r^{-(n/2-2)}) \,,\,
  \delta g_{ur}\,=\,O(r^{-(n-2)})\,,\,\delta g_{IJ}\,=\,O(r^{-(n/2-3)}) \label{AS2},
  \end{gather}
%
%==================================%
where $\delta g_{ab}\equiv\mathcal{L}_{\xi} g_{ab} = 2\hat{\nabla}_{(a}\xi_{b)}$ and $\xi$ is the generator of asymptotic symmetry. 
In the following we consider the asymptotic symmetry in $n>4$ dimensional spacetimes.   
  
The condition of Eq.~(\ref{AS1}) comes from the definition of the Bondi coordinates and 
the explicit forms are 
%============<Equation>=============%
%
\begin{equation}
   \delta g_{rr} = \mathcal L_\xi g_{rr} = 
    - 2 e^B (\xi^u)' = 0,  
\end{equation}
%
%==================================%
%============<Equation>=============%
%
\begin{equation}
   \delta g_{rI} = \mathcal L_\xi g_{rI} =
    - e^B D_I \xi^u + \gamma_{IJ}C^J(\xi^u)'
    + \gamma_{IJ}(\xi^J)' = 0,
\end{equation}
%
%==================================%
%============<Equation>=============%
%
\begin{equation}
   \gamma^{IJ}\delta g_{IJ} = \gamma^{IJ}\mathcal L_\xi g_{IJ} =
    \xi^r (\log\gamma)' + \xi^u \dot{(\log\gamma)}
    + 2 D_I \xi^I + 2 C^I D_I\xi^u = 0,
\end{equation}
%
%==================================%
where $\gamma \equiv \det\gamma_{IJ}$. 
Then, using $\gamma_{IJ} = r^2 h_{IJ}$ and the gauge condition of Eq.~(\ref{gauge}), we can obtain $\xi^{a}$ satisfying the above equations as
%============<Equation>=============%
%
\begin{equation}
   \xi^u = f (u, x^I),
\end{equation}
%
%==================================%
%============<Equation>=============%
%
\begin{equation}
   \xi^I = f^I (u, x^I) + \int\mathrm dr \frac{e^B}{r^2} h^{IJ}\nabla_J f (u, x^I),
\end{equation}
%
%==================================%
%============<Equation>=============%
%
%\begin{equation}
%   \begin{aligned}
%    \xi^r =& - \frac{r}{n-2}
%    (C^I D_I f + D_I f^I
%    + D_I\int\mathrm dr e^B D^I f)\\
%    =& - \frac{r}{n-2}
%    (C^I \nabla_I f + \nabla_I \xi^I)
%   \end{aligned}.
%\end{equation} 
\begin{equation}
 \xi^r = - \frac{r}{n-2}(C^I \nabla_I f + \nabla_I \xi^I).
\end{equation} 
%
%==================================%
For later convenience, we write down the asymptotic behavior of $\xi$ near null infinity as 
%============<Equation>=============%
%
\begin{equation}
   \xi_{r}\,=\,-f(u,x^{I}) +O(r^{-(n-2)}) ,
\end{equation}
\begin{equation}
 \begin{split}
   \xi_{I}\,=\,&r^{2}\omega_{IJ}f^{J}(u,x^{K}) -r\nabla_{I}f(u,x^{K})\\
           &+\sum_{k=0}^{k<n/2-2}\left(
  r h^{(k+1)}_{IJ}f^{J}
  %\frac{h^{(k+1)}_{IJ}f^{J}}{r}
  +fC^{(k+1)}_{I}
           -\frac{n+2k-2}{n+2k}h^{(k+1)}{}^{~J}_{I}\nabla_{J}f
           \right)r^{-(n/2+k-2)}
	   +O(r^{-(n-3)}),
 \end{split}
\end{equation}
\begin{equation}
   \begin{aligned}
    \xi_{u}\,=\,\frac{r}{n-2}\Bigg[\nabla_{I}f^{I}-\frac{1}{r}(\nabla^{2}f+(n-2)f)
           &+&(n-2)\sum_{k=0}^{k<n/2-2}\left( f^{I}C^{(k+1)}_{I}
    - \frac{f A^{(k+1)}}{r} - \frac{n+2k-4}{n+2k-2}
           \frac{C^{(k+1)}_{I}\nabla^{I}f}{r}\right. \\
           &&\left.
           +\frac{2}{n+2k}
           \frac{h^{(k+1)}_{IJ}\nabla^{I}\nabla^{J}f}{(n-2)r} 
           \right) r^{-(n/2+k-1)} \Bigg] +O(r^{-(n-3)}).
   \end{aligned}
\end{equation}
%
%==================================% 
Next let us consider the boundary conditions of Eq.~(\ref{AS2}).
The each components of metric variations are
%============<Equation>=============%
%
\begin{eqnarray}
\delta g_{uu}&=&\frac{2r}{n-2}\frac{\partial}{\partial u}\nabla_{I}f^{I}
               -\frac{2}{n-2}\frac{\partial}{\partial u}(\nabla^{2}f+(n-2)f)
	       + 2C^{(1)}_{I}\partial_{u}f^{I} r^{-(n/2-2)}
	       +O(r^{-(n/2-1)}), 
\label{guu}\\
\delta g_{ur}&=&\frac{1}{n-2}[\nabla_{I}f^{I}-(n-2)\partial_{u}f]
                -\sum_{k=0}^{k<n/2-2}\frac{n+2k-2}{(n-2)(n+2k)}h^{(k+1)}_{IJ}
                \nabla^{I}\nabla^{J}f r^{-(n/2+k)} +O(r^{-(n-2)}), 
\label{gur}\\
\delta g_{uI}&=&r^2 \partial_{u}f_I
               +\frac{r}{n-2}\partial_{I}[\nabla_{J}f^{J}-(n-2)\partial_{u}f]
               -\frac{1}{n-2}\partial_{I}[\nabla^{2}f+(n-2)f]\nonumber\\
	       &&+h^{(1)}_{IJ}\partial_{u}f^J r^{-(n/2-3)}
	       +O(r^{-(n/2-2)}), 
\label{guI}\\
\delta g_{IJ}&=&2r^{2}\Bigg[\nabla_{(I}f_{J)}-\frac{\nabla_{K}f^{K}}{n-2}
                \omega_{IJ}\Bigg] -2r\Bigg[\nabla_{I}\nabla_{J}f 
                -\frac{\nabla^{2}f}{n-2}\omega_{IJ}
                \Bigg] +O(r^{-(n/2-3)}).
\label{gIJ}
\end{eqnarray}
%
%==================================%
To satisfy the boundary conditions of Eq.~(\ref{AS2}) for these equations, we will find that $f$ and $f^{I}$
should satisfy 
%============<Equation>=============%
%
\begin{gather}
\partial_{u}f^{I}\,=\,0 ,\label{fI} \\
\nabla_{I}f_{J}+\nabla_{J}f_{I}\,=\,\frac{2\nabla_K f^K}{n-2}\omega_{IJ},\quad
\nabla_I f^I = (n-2)\frac{\partial f}{\partial u},
\label{fI2} \\
\nabla_{I}\nabla_{J}f\,=\,\frac{\nabla^{2}f}{n-2}\omega_{IJ}\label{f}.
\end{gather}
%
%==================================%
%where we used $h_{IJ}^{(0)}=\omega_{IJ}$. 
%Noe that the condition (\ref{f}) is not required for four dimensions. 
%Note that we consider only the longitudinal mode of $f^{I}$ because the transverse part of $f^{I}$ which 
%satisfies Eq.~(\ref{fI2}) corresponds to a Killing vector on $S^{n-2}$ as seen soon later. 
%Thus the transformations generated by the transverse part of $f^{I}$ are trivial and 
%we consider only longitudinal part of $f^{I}$ which generate non-trivial transformations. 

Integrating the trace part of Eq.~(\ref{fI2}), we can obtain
%============<Equation>=============%
%
\begin{eqnarray}
f\,=\, \frac{F(x^{I})}{n-2}u+\alpha (x^{I}), \label{f2}
\end{eqnarray}
%
%==================================%
where $F\,\equiv\,\nabla_{I}f^{I}$ and $\alpha(x^{I})$ is an integration function on $S^{n-2}$. 
%Here we note that the transverse part of $f^{(\text{tra})I}$ implies 
%$F=\nabla^{I}f^{\text{(tra)}}_{I}=0$ and this corresponds to the vanishing of the right-hand 
%side in Eq.~(\ref{fI2}). This means that $f^{(\text{tra})I}$ corresponds to the 
%Killing vector on $S^{n-2}$. Therefore, as said before, we focused on the longitudinal parts of 
%$f^I$. 
Here we can show from Eqs.~(\ref{f}) and (\ref{f2}) that $F$ satisfies
%============<Equation>=============%
%
\begin{eqnarray}
\nabla_{I}\nabla_{J}F\,=\,\frac{1}{n-2}\omega_{IJ}\nabla^{2}F \label{F1},
\end{eqnarray}
%
%==================================%
and 
also contracting Eq.~(\ref{fI2}) with $\nabla^I\nabla^J$ we have 
%============<Equation>=============%
%
\begin{eqnarray}
\nabla^{2}F+(n-2)F\,=\,0 \label{F2}.
\end{eqnarray}
%
%==================================%
%See Appendix~\ref{app:derivations} for the details of the calculations. 
The general solutions to these equations for $F$ are the $l=1$ modes 
of the scalar harmonics on $S^{n-2}$. Next from Eqs.~(\ref{f}) and (\ref{f2}) we can see that
%============<Equation>=============%
%
\begin{eqnarray}
\nabla_{I}\nabla_{J}\alpha \,=\,\frac{1}{n-2}\omega_{IJ} \nabla^{2}\alpha.
\end{eqnarray}
%
%==================================%
should hold in $n>4$ dimensions. The general solutions to this equations are $l=0$ and $l=1$ 
modes of the scalar harmonics on $S^{n-2}$. 
%In four dimensions, there is no restriction on $\alpha$, that is, 
%$\alpha$ is an arbitrary function on $S^{2}$.  

To be summarized, $f$ can be written as
%============<Equation>=============%
%
\begin{eqnarray}
f\,=\,f_{0} + f_{1}(u,x^{I}),
\end{eqnarray}
%
%==================================%
where $f_{0}$ is a constant and corresponds to the $l=0$ mode in $\alpha$. 
$f_{1}(u,x^{I})$ contains the $l=1$ modes in $F$ and $\alpha$ for $n>4$ dimensions.  
%Since  $\alpha$ is an arbitrary function on $S^{2}$ in four dimensions, 
%$f_{1}(u,x^{I})$ can contain all $l$ modes on $S^{2}$.   
Thus we can show that $f$ satisfies the following equations:
%============<Equation>=============%
%
\begin{eqnarray}
\nabla_{I}(\nabla^{2}f+(n-2)f)\,=\,0
\,,\,
\partial_{u}(\nabla^{2}f+(n-2)f)\,=\,0, \label{f3}
\end{eqnarray}
%
%==================================%
in $n>4$ dimensions. 
%For the comparison we note that the former condition, which come from Eq.~(\ref{guI}) 
%is not required in four dimensions. This is 
%because the metric variation (\ref{guI}) can satisfy the boundary condition (\ref{AS2})
%without the former condition. 
%In four dimensions only $\partial_{u}(\nabla^{2}f+(n-2)f)\,=\,0$ holds 
%and the condition $\nabla_{I}(\nabla^{2}f+(n-2)f)\,=\,0$ is not needed to satisfy the boundary 
%conditions. 
In addition, since $\nabla_{I}\nabla_{J}f \propto \omega_{IJ}$ and $h^{(k+1)}_{IJ}$ are traceless
for $k<n/2-1$, the gauge condition of Eq.~(\ref{gauge}) implies that 
$h^{(k+1)}_{IJ}\nabla^{I}\nabla^{J}f$ vanishes for $k<n/2-2$ in Eq.~(\ref{gur}).
%Using the above equations, we see that Eq.~(\ref{AS2}) become
%%============<Equation>=============%
%%
%\begin{eqnarray}
%\delta g_{uu}&=& O(r^{-(n/2-1)}), \\
%\delta g_{ur}&=&-\sum_{k=0}^{k<n/2-2}\frac{n+2k-2}{(n-2)(n+2k)}h^{(k+1)}_{IJ}
%                \nabla^{I}\nabla^{J}f r^{-(n/2+k)} +O(r^{-(n-2)}), 
%\label{gur2}\\
%\delta g_{uI}&=&O(r^{-(n/2-2)}), 
%\label{guI2}\\
%\delta g_{IJ}&=&2r^{2}\Bigg[\nabla_{(I}f_{J)}-\frac{\nabla^{I}f_{I}}{n-2}
%                \omega_{IJ}\Bigg] +O(r^{-(n/2-3)}).
%\label{gIJ2}
%\end{eqnarray}
%%
%%==================================%
%Since $\nabla_{I}\nabla_{J}f \propto \omega_{IJ}$ and $h^{(k+1)}_{IJ}$ are traceless
%for $k<n/2-1$, the gauge condition of Eq.~(\ref{gauge}) implies that 
%$h^{(k+1)}_{IJ}\nabla^{I}\nabla^{J}f$ vanishes for $k<n/2-2$ in Eq.~(\ref{gur2}). Noting that the condition (\ref{fI2}) can be rewritten
%as
%%============<Equation>=============%
%%
%\begin{eqnarray}
%\nabla_{(I}f_{J)}-\frac{\nabla^{I}f_{I}}{n-2}\omega_{IJ}=0,
%\end{eqnarray}
%%
%%==================================%
%we can show that $\delta g_{uI}=O(r^{-(n/2-3)})$. 
As a consequence, we could confirm that 
the transformations satisfying Eqs.~(\ref{fI}), (\ref{fI2}) and (\ref{f}) 
keep the boundary conditions (\ref{AS2}). 

It is worth noting that Eq.~(\ref{fI2}) gives another condition for 
$f^{(\mathrm{tra})I}$ which is the transverse part of $f^I$, namely 
satisfying $\nabla_I f^{(\mathrm{tra})I} = 0$. 
We find that Eq.~(\ref{fI2}) corresponds to the 
Killing equation 
$\nabla_{I} f^{(\mathrm{tra})}_{J} + \nabla_{J}f^{(\mathrm{tra})}_{I}=0$ 
on $S^{n-2}$ because of the transverse condition. 
This means that $f^{(\mathrm{tra})I}$ is the Killing vector on 
$S^{n-2}$. 
Therefore, the transformations generated by the transverse part of 
$f^{I}$ are trivial and we could focus on only the longitudinal part of $f^{I}$ 
which generates nontrivial transformations.

Here we give the short summary. We could show that the asymptotic symmetry is generated 
by $f$ and $f^{I}$ satisfying 
Eqs.~(\ref{fI}), (\ref{fI2}) and (\ref{f}). The parts of $f$, which are not proportional 
to $u$, generates a translation group. 
$f^{I}$ generates the Lorentz group. Then the asymptotic symmetry at null infinity is the 
Poincar\'e group.

Before closing this subsection, we have a comment on four dimensional cases for the comparison. 
In four dimensional cases, the boundary conditions to be held are 
%============<Equation>=============%
%
\begin{eqnarray}
\delta g_{uu}\,=\,O(r^{-1})\,,\,
\delta g_{ur}\,=\,O(r^{-2})\,,\,
\delta g_{uI}\,=\,O(1)\,,\,
\delta g_{IJ}\,=\,O(r) \label{AS3}.
\end{eqnarray}
%
%==================================%
Then, if $f$ and $f^{I}$ satisfy Eqs. (\ref{fI}) and (\ref{fI2}), the transformations keep the 
above boundary conditions. Note that the condition (\ref{f}) is not required  
because the second term in the right-hand side in Eq.~(\ref{gIJ}) already satisfies the boundary 
conditions and has no additional restriction to $f$ in four dimensions. 
Therefore there is no restriction on $\alpha$ where $f=F(x^{I})u/2+\alpha$.
Hence $\alpha$ is an arbitrary functions on $S^{2}$ in four dimensions while in $n>4$ dimensions 
$\alpha$ should be $l=0$ or $l=1$ mode. The former condition 
$\nabla_{I}(\nabla^{2}f+(n-2)f)\,=\,0$ in Eq.~(\ref{f3}), which comes from 
Eq.~(\ref{guI}), does not hold in four dimension because $\alpha$ can have $l>1$ modes. 
However since the third term in the right-hand side in Eq.~(\ref{guI}) already satisfies the 
boundary conditions (\ref{AS3}) in four dimensions, that condition is not required. The $l>1$ modes of $\alpha$ correspond to the generators of 
the so-called supertranslation group. Thus the asymptotic symmetry is the semi-direct group of 
supertranslation and Lorentz group in four dimensions rather than the 
Poincar\'e group.

\subsection{Poincar\'e covariance}
  
Next we shall confirm the Poincar\'e covariance of the Bondi mass and momentum. 
Since the asymptotic symmetry is the Poincar\'e group, we expected that the 
Bondi mass and momentum should be transformed covariantly under the action of 
its Poincar\'e group. In practice, under the translation of $f=\alpha(x^{I})$ and $f^{I}=0$, 
the Bondi energy-momentum is invariant, that is, 
%============<Equation>=============%
%
\begin{eqnarray}
    M^{a}_{\text{Bondi}}\rightarrow M^{a}_{\text{Bondi}}.
\end{eqnarray}
%
%==================================%
However, since we consider dynamical spacetimes, it is easy to expect the contribution from 
gravitational waves under translation $u\rightarrow u-\alpha$. Thus the Bondi energy-momentum 
$M^{a}_{\text{Bondi}}$ should be transformed under the translation as
%============<Equation>=============%
%
\begin{eqnarray}
   M^{a}_{\text{Bondi}}(u)\rightarrow 
    M^{a}_{\text{Bondi}}(u) + \mathcal L_\xi M^{a}_{\text{Bondi}}
                    =M^{a}_{\text{Bondi}}(u) +\alpha\frac{d}{du}M^{a}_{\text{Bondi}}(u),
\end{eqnarray}
%
%==================================% 
where the second term in the right-hand side represents the effect of gravitational radiations. 
Let us look at the details. 
 
For the translations, the generator $\xi_{a}$ becomes
%============<Equation>=============%
%
\begin{equation}
   \xi_{r}\,=\,-\alpha +O(r^{-(n-2)}),
  \end{equation}
  \begin{equation}
   \xi_{I}\,=\,-r\nabla_{I}\alpha +\sum_{k=0}^{k<n/2-2}\Bigg[ \alpha C^{(k+1)}_{I}
   -\frac{n+2k-2}{n+2k}h^{(k+1)}_{IJ}\nabla^{J}\alpha \Bigg] r^{-(n/2+k-2)}+O(r^{-(n-3)}),
  \end{equation}
  \begin{equation}
   \xi_{u}\,=\,-\sum_{k=0}^{k<n/2-2}\Bigg[\frac{n+2k-4}{n+2k-2}C^{(k+1)}_{I}\nabla^{I}\alpha
   +\alpha A^{(k+1)}\Bigg] r^{-(n/2+k-1)} +O(r^{-(n-3)}).
\end{equation}
%
%==================================%
$\delta g_{uu}$ can be computed as
%============<Equation>=============%
%
\begin{eqnarray}
   \delta g_{uu}&=& 2\hat{\nabla}_{u}\xi_{u} \notag \\
                &=& \sum_{k=0}^{k=n/2-2}\delta g^{(k+1)}_{uu}r^{-(n/2+k-1)}+O(r^{-(n-5/2)}),  
\end{eqnarray}
%
%==================================%
where 
%============<Equation>=============%
%
\begin{eqnarray}
   \delta g^{(k+1)}_{uu}\,=\,\frac{4}{n+2k-2}\partial_{u}C^{(k+1)}_{I}\nabla^{I}\alpha
    -\alpha \partial_{u}A^{(k+1)} -\frac{n+2k-4}{2}\alpha A^{(k)}+\nabla^{I}\alpha
    \nabla_{I}A^{(k)}.
\end{eqnarray}
%
%==================================%
In particular, for $k=n/2-2$
%============<Equation>=============%
%
\begin{eqnarray}
   \delta g^{(n/2-1)}_{uu}&=& \delta m \notag \\
   &=& \alpha \partial_{u}m+\frac{2}{n-3}\nabla^{I}\alpha \partial_{u}C^{(n/2-1)}_{I}
   -\alpha (n-4)A^{(n/2-2)}+\nabla^{I}\alpha\nabla_{I}A^{(n/2-2)}.
\end{eqnarray}
%
%==================================%
By using Eqs.~(\ref{Asol}), (\ref{A-evo})
and (\ref{const}), we can rewrite $\delta g^{(k+1)}_{uu}$ as
%============<Equation>=============%
%
\begin{eqnarray}
   \delta g^{(k+1)}_{uu}\,=\,\frac{2}{n+2k}[\nabla^{2}(\alpha A^{(k)})+(n-2)\alpha A^{(k)}]
   &+&\frac{4}{(n+2k)(n-2k-2)}\nabla^{I}\nabla^{J}(\alpha\partial_{u}h^{(k+1)}_{IJ})\notag \\
   &-&\frac{2(n+2k-6)}{(n+2k)(n-2k-2)}[\nabla^{I}\nabla^{J}(\nabla_{I}\alpha C^{(k)}_{J})
    +C^{(k)}_{I}\nabla^{I}\alpha ],
\end{eqnarray}
%
%==================================%
for $0\leq k<n/2-2 $ and
%============<Equation>=============%
%
\begin{eqnarray}
    \delta m &=&\,\alpha\partial_{u}m+\frac{2}{n-3}\nabla^{I}\alpha\partial_{u}C_{I}^{(n/2-1)}
    -(n-4)\alpha A^{(n/2-2)}+\nabla^{I}\alpha \nabla_{I}A^{(n/2-2)} \notag \\ 
    &=&-\frac{\alpha}{2(n-2)}\dot{h}^{(1)}_{IJ}\dot{h}^{(1)IJ}+\frac{2}{n-2}\nabla^{I}\nabla^{J}
    (\alpha\partial_{u}h^{(n/2-1)}_{IJ})+\frac{1}{n-2}[\nabla^{2}(\alpha A^{(n/2-2)})+(n-2)\alpha 
    A^{(n/2-2)}] \notag\\ 
    && ~~~~~~~~~~~~~~~~~~~~~~~~~~~~~~~~~~~~~~~~~~~~~~~~~~
    -\frac{n-5}{n-2}[\nabla^{I}\nabla^{J}(\nabla_{I}\alpha C_{J}^{(n/2-2)})
    +C^{(n/2-2)}_{I}\nabla^{I}\alpha ],
   \end{eqnarray}
%
%==================================%
for $k=n/2-2$. 
We can show that 
%============<Equation>=============%
%
\begin{eqnarray}
\int_{S^{n-2}}\delta g^{(k+1)} d\Omega \,=\,0\,,\,~~{\rm and}~~
\int_{S^{n-2}}\hat{x}^{i}\delta g^{(k+1)}_{uu}d\Omega\,=\,0 \label{B-finite},
\end{eqnarray}
%
%==================================%
for $k<n/2-2$ and
%============<Equation>=============%
%
\begin{eqnarray}
\int_{S^{n-2}}\delta md\Omega\,=\,-\frac{1}{2(n-2)}\int_{S^{n-2}}\alpha
   \dot{h}^{(1)}_{IJ}\dot{h}^{(1)IJ}d\Omega,
   \label{B-trans} 
\end{eqnarray}
%
%==================================%
for $k=n/2-2$. See Appendix~\ref{app:derivations} for the details of the calculations.

Equation (\ref{B-finite}) implies that translations 
$u \rightarrow u-\alpha$ preserve the finiteness of the Bondi energy-momentum. 
Equation (\ref{B-trans}) can be rewritten as 
%============<Equation>=============%
%
\begin{eqnarray}
   M^{a}_{\text{Bondi}}\rightarrow M^{a}_{\text{Bondi}} 
   +\alpha\frac{d}{du}M^{a}_{\text{Bondi}},
    %=M^{a}_{\text{Bondi}}+\mathcal{L}_{\xi}M^{a}_{\text{Bondi}},
\end{eqnarray}
%
%==================================%
where $dM_{\text{Bondi}}/du$ is given by Eq.~(\ref{BML}). Thus the Bondi mass in our definition 
has the Poincar\'e covariance under the asymptotic symmetry.

\section{summary and outlook}
\label{sec:summary}

In this paper, we have investigated the asymptotic structure at null infinity in $n$-dimensional 
spacetimes using the Bondi coordinates. Asymptotic flatness is defined by the asymptotic behavior 
of gravitational fields at null infinity. These boundary conditions are determined by solving 
the Einstein equations. Although the Bondi mass seems to diverge in the conformal method, 
we can show its finiteness from the Einstein equations in the Bondi coordinates. 
And we can show that asymptotic symmetry at null infinity should be the Poincar\'e group 
and the Bondi energy-momentum is transformed covariantly under the Poincar\'e group by using the 
Einstein equations. 
These results are same with those in \cite{Hollands:2003ie, Hollands:2003xp} for even 
dimensions. Note that the conditions for 
asymptotic flatness in \cite{Hollands:2003ie, Hollands:2003xp} come 
from the stability of weak asymptotic simplicity \cite{Geroch:1978ur} .
On the other hands, our definition of asymptotic flatness comes from the behavior of perturbations 
around the Minkowski spacetime. In general, these two definitions may differ. 
The Bondi mass will diverge unless our boundary conditions at null infinity 
are not satisfied. In this sense we would expect that our definition guarantees the stability 
of weak asymptotic simplicity at null infinity. Nevertherless, it is nice to show that our 
definition is generic enough regardless of the Minkowski spacetime as 
Refs. \cite{Hollands:2003ie, Hollands:2003xp}.

 As our future work we will be able to consider angular momentum at null infinity in 
$n$-dimensional spacetimes. Since asymptotic symmetry at null infinity is 
the Poincar\'e group without supertranslations in higher dimensions, we can define the 
angular momentum. Indeed, we can define the angular momentum and show its Poincar\'e 
covariance in five dimensions 
\cite{Tanabe:2010rm}.

\begin{acknowledgments}
KT is supported by JSPS Grant-Aid for Scientific Research (No.$~21$-$2105$).
SK is the Yukawa Fellow and this work is partially supported by Yukawa Memorial Foundation.
TS is partially supported by Grant-Aid for Scientific Research from
Ministry of Education, Science,
Sports and Culture of Japan (Nos.~21244033,~21111006,~20540258 and
19GS0219).
This work is also supported by the Grant-in-Aid for the Global 
COE Program ``The Next Generation of Physics, Spun from Universality 
and Emergence'' from the Ministry of Education, Culture, Sports, Science 
and Technology (MEXT) of Japan. 
\end{acknowledgments}

\appendix
 
 \section{$((n-1)+1)$-decomposition}
 \label{app:decomposition}
 
The $n$-dimensional metric can be written as 
\begin{equation}
  g_{ab} = \epsilon n_a n_b + \gamma_{ab},
\end{equation}
where $\gamma_{ab}$ is $(n-1)$-dimensional induced metric and $n^a$ is the 
unit normal vector, which is normalized by $n_a n^a = \epsilon$. 
Note that $\epsilon$ takes $+1$ or $-1$ which means the normal vector 
is spacelike or timelike, respectively. 

We define the extrinsic curvature as 
\begin{equation}
  K_{ab} = \frac{1}{2}\mathcal L_n \gamma_{ab}.
\end{equation}

Because $n_a$ is the normal vector to the $(n-1)$-dimensional hypersurface, 
it can be written as $n_a = \epsilon N \nabla_a \Omega$ where $\Omega$ 
is a function which describes the hypersurface by 
$\Omega = \text{const.}$ and $N$ is so-called lapse function.
Then, the Riemann tensor becomes 
\begin{equation}
  R_{efgh}\gamma_a{}^e\gamma_b{}^f\gamma_c{}^g\gamma_d{}^h
   = {}^{(\gamma)}R_{abcd} - \epsilon K_{ac}K_{bd} + \epsilon K_{ad}K_{bc},
\end{equation}
\begin{equation}
  R_{efgd}\gamma_a{}^e\gamma_b{}^f\gamma_c{}^g n^d
   = D_aK_{bc} - D_bK_{ac},
\end{equation}
\begin{equation}
  R_{acbd}n^cn^d = - \mathcal L_n K_{ab} + K_{ac}K_b{}^c 
   - \epsilon \frac{1}{N}D_aD_b N,
\end{equation}
where $D_a$ denotes the covariant derivative with respect to $\gamma_{ab}$.
Note that we have used $n^a\nabla_a n_b = - \epsilon D_b\log N$.
 %where we have defined the normal vector as 
 %$n_a = \epsilon N \nabla_a \Omega$. 

The Ricci tensor becomes 
\begin{equation}
  R_{ab}n^an^b = - \mathcal L_n K - K_{ab}K^{ab}-\epsilon \frac{1}{N}D^2N,
\end{equation}
\begin{equation}
  R_{ac}n^a\gamma_b{}^c = D^a K_{ab} - D_b K,
\end{equation}
\begin{equation}
  R_{cd}\gamma_a{}^c\gamma_b{}^d = 
   {}^{(\gamma)}R_{ab} - \epsilon\mathcal L_n K_{ab}
   - \epsilon KK_{ab} + 2\epsilon K_{ac}K_b{}^c
   - \frac{1}{N}D_aD_bN
\end{equation}

The Ricci scalar becomes 
\begin{equation}
  \begin{aligned}
   R =& {}^{(\gamma)}R - 2\epsilon \mathcal L_n K - \epsilon K^2
   - \epsilon K_{ab}K^{ab} - \frac{2}{N}D^2N\\
   =& {}^{(\gamma)}R + \epsilon K^2
   - \epsilon K_{ab}K^{ab} - \frac{2}{N}D^2N - 2\epsilon\nabla_a(Kn^a)
  \end{aligned}
\end{equation}

The each components of the Einstein tensor are given by 
\begin{equation}
  G_{ab}n^an^b = \frac{1}{2}
   (- \epsilon {}^{(\gamma)}R + K^2 - K_{ab}K^{ab}),
 \end{equation}
 \begin{equation}
  G_{ac}n^a\gamma_b{}^c = D^a K_{ab} - D_b K,
 \end{equation}
 \begin{equation}
  \begin{aligned}
   G_{cd}\gamma_a{}^c\gamma_b{}^d =& 
   {}^{(\gamma)}G_{ab}
   - \epsilon KK_{ab} + 2\epsilon K_{ac}K_b{}^c
   + \frac{\epsilon}{2}\gamma_{ab}(K_{cd}K^{cd} + K^2)
   \\
   & - \epsilon\mathcal L_n K_{ab} + \epsilon\gamma_{ab}\mathcal L_n K
   - \frac{1}{N}D_aD_bN + \frac{1}{N}\gamma_{ab}D^2N 
  \end{aligned}
 \end{equation}

\section{Derivations of (\ref{B-finite}) and (\ref{B-trans}).}
\label{app:derivations}

We will show Eq.~(\ref{B-finite}) and (\ref{B-trans}). 
At first we show the former equation in Eq.~(\ref{B-finite}).
Since the integrations of the total derivative terms vanish, we can obtain 
%============<Equation>=============%
%
\begin{eqnarray}
\int_{S^{n-2}}\delta g^{(k+1)}_{uu}d\Omega\,=\,\int_{S^{n-2}}\Bigg[
\frac{2(n-2)}{n+2k}\alpha A^{(k)}-\frac{2(n+2k-6)}{(n+2k)(n-2k-2)}\nabla^{I}\alpha C^{(k)}_{I}
\Bigg]d\Omega.
\end{eqnarray}
%
%==================================%
Using Eqs.~(\ref{Csol}) and (\ref{Asol}), we can see that
%============<Equation>=============%
%
\begin{eqnarray}
\int_{S^{n-2}}C^{(k)I}\nabla_{I}\,\alpha\, d\Omega &=& 
\frac{2(n+2k-4)}{(n+2k-2)(n-2k)}\int_{S^{n-2}}\nabla_{J}h^{(k)IJ}\nabla_{I}\alpha d\Omega 
\notag \\
&=& -\frac{2(n+2k-4)}{(n+2k-2)(n-2k)}\int_{S^{n-2}}h^{(k)IJ}\nabla_{I}\nabla_{J}\alpha d\Omega 
\notag \\
&=&0, \label{Cint}
\end{eqnarray}
%
%==================================%
and
%============<Equation>=============%
%
\begin{eqnarray}
\int_{S^{n-2}}\alpha A^{(k)}d\Omega &=&
-\frac{4(n+2k-6)}{(n+2k-2)(n-2k)(n-2k-2)}\int_{S^{n-2}}\alpha\nabla^{I}\nabla^{J}h_{IJ}^{(k)}d\Omega
\notag \\
&=&-\frac{4(n+2k-6)}{(n+2k-2)(n-2k)(n-2k-2)}\int_{S^{n-2}}h^{(k)}_{IJ}\nabla^{I}\nabla^{J}\alpha d\Omega 
\notag \\
&=&0, \label{Aint}
\end{eqnarray}
%
%==================================%
where we used the fact that $h^{(k)}_{IJ}$ are traceless for $k<n/2$. 
Then we can show 
%============<Equation>=============%
%
\begin{eqnarray}
\int_{S^{n-2}}\delta g^{(k+1)}_{uu} d\Omega \,=\,0.
\end{eqnarray}
%
%==================================%
This is the former one in Eq.~(\ref{B-finite}). 

Next we show the latter one in Eq.~(\ref{B-finite}). For this we can see 
%============<Equation>=============%
%
\begin{eqnarray}
\int_{S^{n-2}}\hat{x}^{i}[\nabla^{2}(\alpha A^{(k)})+(n-2)\alpha A^{(k)}]d\Omega
\,=\,\int_{S^{n-2}}\alpha A^{(k)}[\nabla^{2}\hat{x}^{i} +(n-2)\hat{x}^{i}]d\Omega
\,=\,0,
\end{eqnarray}
%
%==================================%
%============<Equation>=============%
%
\begin{eqnarray}
\int_{S^{n-2}}\hat{x}^{i}\nabla_{I}\nabla_{J}(\alpha h^{(k)}_{IJ})d\Omega 
\,=\,\int_{S^{n-2}}\alpha h^{(k)}_{IJ}\nabla^{I}\nabla^{J}\hat{x}^{i}d\Omega
\,=\,0,
\end{eqnarray}
%
%==================================%
and
%============<Equation>=============%
%
\begin{eqnarray}
\int_{S^{n-2}}\hat{x}^{i}[\nabla^{I}\nabla^{J}(\nabla_{I}\alpha C^{(k)}_{J})
+\nabla^{I}\alpha C^{(k)}_{I}]d\Omega\,=\,\int_{S^{n-2}}
\nabla_{I}\alpha C^{(k)}_{J}
[
\nabla^{I}\nabla^{J}\hat{x}^{i}+\omega^{IJ}\hat{x}^{i}
]d\Omega \,=\,0
\end{eqnarray}
%
%==================================%
hold. In the aboves we used the tracelessness of $h^{(k)}_{IJ}$. 
Using of them, we can show 
%============<Equation>=============%
%
\begin{eqnarray}
\int_{S^{n-2}}\hat{x}^{i}\delta g^{(k+1)}_{uu}d\Omega\,=\,0 .
\end{eqnarray}
%
%==================================%
This is the latter one in Eq.~(\ref{B-finite}). 

Finally we show Eq.~(\ref{B-trans}).
Since the integrations on $S^{n-2}$ of the total derivative terms vanish,  
%============<Equation>=============%
%
\begin{eqnarray}
\int_{S^{n-2}} \delta m d\Omega &=&-\frac{1}{2(n-2)}\int_{S^{n-2}}\alpha
   \dot{h}^{(1)}_{IJ}\dot{h}^{(1)IJ}d\Omega +
   \int_{S^{n-2}}\Bigg[
   \alpha A^{(n/2-2)}-\frac{n-5}{n-2}C^{(n/2-2)I}\nabla_{I}\alpha
   \Bigg] d\Omega 
   \notag \\
   &=&-\frac{1}{2(n-2)}\int_{S^{n-2}}\alpha
   \dot{h}^{(1)}_{IJ}\dot{h}^{(1)IJ}d\Omega,
\end{eqnarray}
%
%==================================%
where we used Eqs.~(\ref{Cint}) and (\ref{Aint}) from the first to second line. 
This is Eq.~(\ref{B-trans}).

\end{document}